\documentclass[letterpaper,12pt,peerreviewca,draftcls]{IEEEtran}
\usepackage{csm16}
\usepackage{graphicx,xcolor}
\usepackage{amsmath,amsthm,amssymb}
\usepackage{cite}

\title{Disentangling Resilience from Robustness: Contextual Dualism, Interactionism, and Game-Theoretic Paradigms}

\author{Quanyan Zhu and Tamer Ba\c{s}ar\thanks{Q. Zhu is with New York University, E-mail: qz494@nyu.edu; and T. Ba\c{s}ar is with the University of Illinois, Urbana-Champaign, E-mail: basar1@illinois.edu}}
\begin{document}

\maketitle
\CSMsetup

\begin{abstract}
 This article explains the distinctions between robustness and resilience in control systems. Resilience confronts a distinct set of challenges,  posing new ones for designing controllers for feedback systems, networks, and machines that prioritize resilience over robustness.  The concept of resilience is explored through a three-stage model, emphasizing the need for a proactive preparation and automated response to elastic events.  A toy model is first used to illustrate the tradeoffs between resilience and robustness. Then, it delves into contextual dualism and interactionism, and introduces game-theoretic paradigms as a unifying framework to consolidate resilience and robustness.  The article concludes by discussing the interplay between robustness and resilience, suggesting that a comprehensive theory of resilience and quantification metrics, and formalization through game-theoretic frameworks are necessary. The exploration extends to system-of-systems resilience and various mechanisms, including the integration of AI techniques and non-technical solutions, like cyber insurance, to achieve comprehensive resilience in control systems. \textcolor{black}{As we approach 2030, the systems and control community is at the opportune moment to lay scientific foundations of resilience by bridging feedback control theory, game theory, and learning theory.  Resilient control systems will enhance overall quality of life, enable the development of a resilient society, and create a societal-scale impact amid global challenges such as climate change, conflicts, and cyber insecurity.}
\end{abstract}

\textcolor{black}{As we gaze into the landscape of 2030, \emph{resilience}  emerges as the cornerstone for addressing these societal-scale challenges that lie ahead. It is central to all engineering systems today grappling with uncertainties arising not only from the physical environment but also from cyber domains and human factors. }
In the cyber domain, recent years have witnessed increasingly sophisticated attacks, such as the SolarWinds breach \cite{willett2023lessons} and ransomware incidents \cite{nicol2021ransomware}. These attacks have disrupted systems and left plant operators ill-prepared and uncertain about how to effectively respond to such disruptive events. Human factors contribute significantly to this challenge as well. Human negligence stands out as a major catalyst for cyberattacks, with phishing, spear-phishing, insider threats, and human vulnerabilities collectively responsible for 95\% of cyberattacks, as reported by IBM \cite{SecurityMagazineRSS_2020}. Furthermore, human errors can lead to unexpected failures in system operations, whether due to deviations from established protocols or inadequate responses to emergent situations.

Control engineers have been used to designing  {\it robust} control systems to withstand and resist anticipated events like measurement disturbances and sensor or actuator faults. However, lessons from recent events have highlighted the limitations of robustness when it comes to events that are hard to prepare for. These events may arise due to a combination of high preparation costs, a low likelihood of occurrence, and the inherent difficulty in their foreseeing and planning for. As engineering systems continue to evolve in complexity, exemplified by the emergence of cyber-physical systems, socio-technical systems, and AI-enabled systems, the frequency of these events is on the rise, and mitigating them is becoming increasingly intricate and costly. Cyberattacks stand out as prime examples of this growing challenge. Attackers have advantages over the defenders in that they have to discover and exploit one single vulnerability to succeed while defenders have to prepare for all the vulnerabilities in order to guarantee the security of a system. It should be clear that such robust defense with the goal of total resistance to adversarial events is cost-prohibitive.

{\it Resilience} has emerged as a new paradigm.
 The primary focus of resilience is no longer solely on creating robust systems that can entirely prevent failures. Instead, the focus has shifted toward mitigating the impact of failures when prevention is impractical. One central part of resilience is the resilience strategy, which is designed to balance among the system performance, criticality, knowledge about the event, and the costs. There is a need for a proactive preparation and automated response to them. Events that were once considered high risk and low frequency, such as terrorist attacks and black swan events,  now occur with greater frequency while still maintaining a high level of risk as the cost associated with launching cyberattacks becomes lower. Consequently, they are becoming ones with higher risk and higher frequency. We call the class of events that require resilience \emph{elastic events}. This class is growing bigger by the day. It often contains events  that are recognized and acknowledged but not precisely quantified or understood (a.k.a. known unknowns) and events that are not identified or anticipated at all (a.k.a. unknown unknowns). In the context of elastic events, resilience becomes imperative, because it is the case where robustness cannot be an option.  Resilience becomes the last resort of defense or control when there are unknown threats. 
 
These distinctions have also been highlighted in the recent literature \cite{annaswamy2016emerging,rieger2019industrial,ishii2022security,sepulchre2023robust}. More specifically, in \cite{sepulchre2023robust}, the author has emphasized that resilience  addresses a unique set of questions, standing apart from robustness. This distinction poses challenges when designing controllers for feedback systems, networks, and machines that prioritize resilience over robustness. This article endeavors to untangle the objectives of resilient control systems from their robust counterparts, providing a nuanced exploration of their interconnection through the lenses of contextual dualism and interactionism. \textcolor{black}{In this article, we refer to disturbances as external influences or perturbations,  including environmental factors, input variations, noise, and faults. Uncertainties refer to the lack of knowledge about future events, outcomes, or states of a system. Uncertainties can arise from various sources, including incomplete information, variability in system parameters, randomness, and unpredictability of external factors. Some events can be uncertainties and disturbances at the same time. However, some uncertainties go beyond disturbances. For example, uncertainties due to the incomplete information of an adversary are often not regarded as simply disturbances as their consequences can be more than perturbations. Norbert Wiener introduced the concept of Manichean and Augustinian devils, providing a nuanced categorization of disturbances and uncertainties into two distinct types. This framework allows the article to introduce} game-theoretic paradigms as a unifying framework to consolidate resilience and robustness, presenting new perspectives to effectively tackle emerging challenges in system and control.

\section{Multi-stage characterization of resilience}

\begin{figure}
    \centering
    \includegraphics[width=0.6\textwidth]{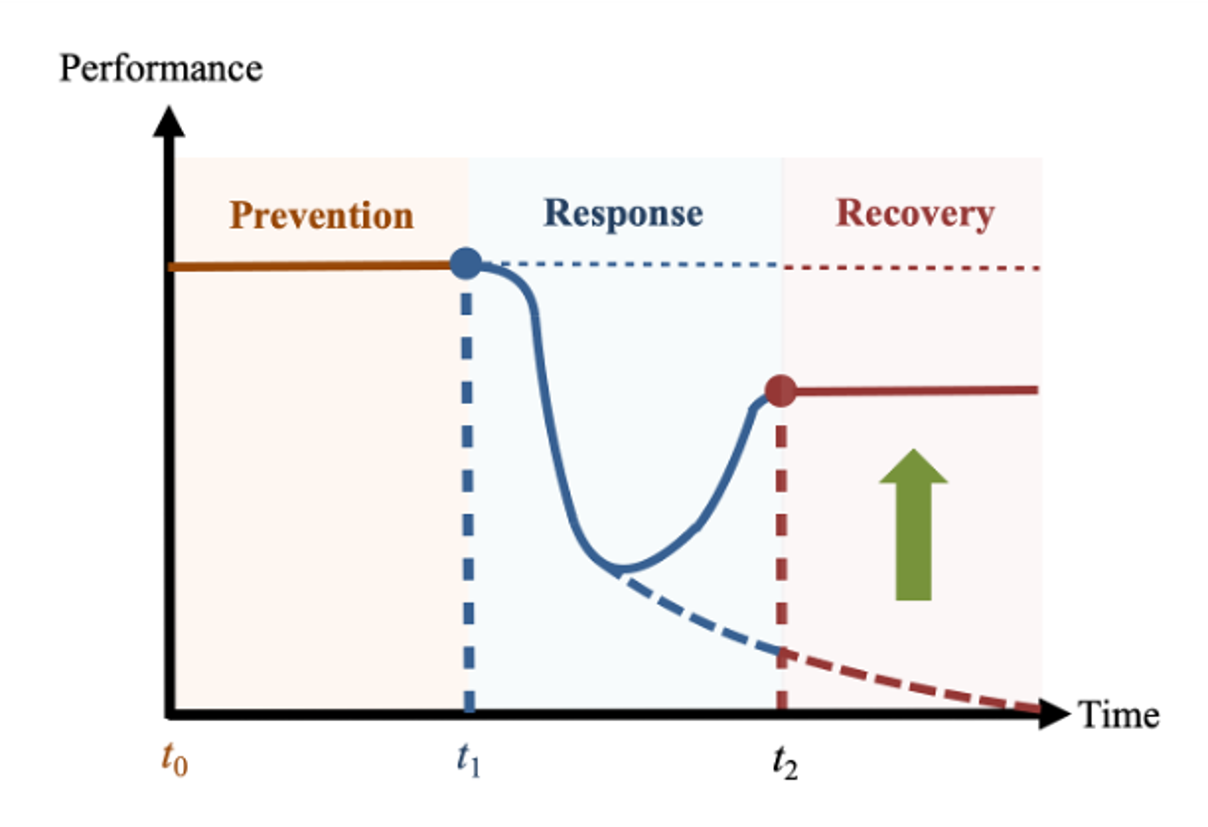}
    \caption{Three-stage illustration of resilience: Resilience focuses on the response to disruptive events and graceful recovery of the system following such disruptions.}
    \label{fig:resilience}
\end{figure}

The concept of resilience is depicted in Figure 1, which breaks down system resilience into three distinct stages. The first stage is the ``{\it ex ante}'' phase, during which the system operates in a controlled manner, striving to prevent disruptive and adversarial events. When a disruptive event occurs and the controlled system cannot withstand the impact of these failures, a degradation in performance ensues. This is when the resilience mechanism comes into play, initiating the ``interim'' stage. The final stage is termed the ``ex post'' phase, wherein efforts are focused on recovering and maintaining the system at an acceptable level of performance following a failure, with this process happening gracefully to the extent possible.

This resilience concept draws an analogy to the human immune response \cite{lee2011self}. Robustness represents the innate defense mechanisms that prevent illness, while resilience involves learning and adapting to combat a virus after infection.

Strategies to attain resilience extend beyond technical solutions and single time scales. Resilience can manifest across multiple temporal dimensions, and it encompasses a spectrum of social and economic strategies. A prime example of an `ex post' strategy is cyber insurance \cite{bohme2010modeling,zhang2017bi,khalili2018designing}. It operates on a different time scale compared to control-oriented solutions, providing a means to mitigate financial losses incurred from cyberattacks or disruptive events days or even months after such incidents occur. By extending coverage, it empowers small and medium-sized enterprises to quickly resume their operations without enduring significant financial burdens.

\section{A toy model for resilience}
In Figure 2, we illustrate a toy model that captures the essence of robustness and resilience. We consider two modes. One is the working mode, denoted by $W$, and the other one is the failure mode, denoted by $F$. The transitions between the two modes are represented by the kernel $\phi_{ij}, i, j \in \{W, F\}$. The transition probability  $\phi_{WF}$ from $W$ to $F$ represents the probability of failure when an event occurs. On the other hand, $\phi_{FW}$ is the transition probability from mode $F$ to $W$. The goal of robustness is to resist exogenous perturbations and disruptions at mode $W$, which could be noises or adversarial behaviors, so that the system stays at mode $W$. The goal of resilience is to recover the system from mode $F$ to mode $W$ so that loss of the system occurring at mode $F$ is minimum.


An ideal robust system is when $\phi_{WF}=0$; i.e., the system has prepared for all possible disruptive events, including elastic and inelastic ones. As a result, the system never fails or enters mode $F$. In this case, resilience is unnecessary. Resilience becomes critical when $\phi_{WF}$ is positive and there is a need to have a sufficiently high $\phi_{FW}$. An ideal resilience system is when $\phi_{FW}=1$; i.e., the system can perfectly recover from any disruptive events. In this case, robustness becomes secondary and its requirement is not stringent since any event eventually will be recovered and go back to normal.

\begin{figure}
    \centering
    \includegraphics[width=0.6\textwidth]{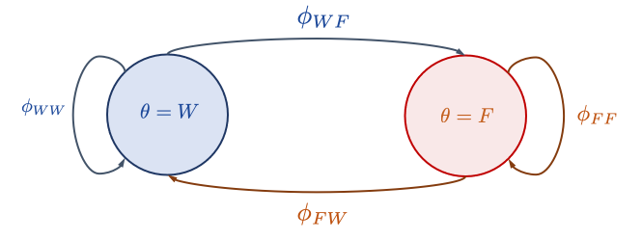}
    \caption{A two-mode toy model to illustrate resilience.}
    \label{fig:toymodel}
\end{figure}

This toy model helps elucidate the fundamental tradeoffs between resilience and robustness. Since ideal robust systems or resilient systems are either non-existent or cost-prohibitive, there are fundamental tradeoffs between the two in the selection of elastic events that aim to be taken care of by resilience and the inelastic events by robustness. The tradeoff will be dependent on the cost associated with the efforts and the impact associated with the events that we are prepared for. More formally, we can consider a set of events, $\mathcal{E}:=\{e_1, e_2, \cdots, e_N\}$. We can partition the set of events into two disjoint sets. One is the set  $\mathcal{E}_F$ of elastic events (that can potentially lead to $F$), and the other one is the set  $\mathcal{E}_W$ of inelastic events (that need to be taken care of at $W$). This partition will determine the transition probabilities and the goal is to find a partition that minimizes the accumulative cost over an operation horizon $T$. 
Timing element is an important element of the resilient control. One main objective of the control law is to make the transition from the failure mode to the working mode take place as quickly as possibly and gracefully. To achieve that, one can design appropriate objective cost functions that capture the time in the failure mode and penalize associated performance losses. 

\textcolor{black}{The self-transitions, $\phi_{WW}$ and $\phi_{FF}$, in the toy examples represent the probability of the systems remaining in the working state or the failure state. The transition is triggered by an elastic event  that leads to the failure state. One important note is that a self-transition is a high-level representation of the system behavior in which the system withstands inelastic events. Phenomenally, the system does not change because it is robust to the inelastic events. If we zoom into the self-transition, there are efforts in designing robust controllers that achieve  phenomenally the transition probability $\phi_{WW}$. Similarly, the self-transition probability $\phi_{FF}$ is a high-level phenomenal consequence of the system staying at $F$ if no resilient control is used to bring the system back to the normal state. It is, in some way, a high-level expression of how resilient the system is in the event of the elastic event. If we zoom into the self-transition $\phi_{FF}$, there are efforts in reducing this probability through resilient control. These two transition probabilities in the toy example align  with the exposition of dualism. On one end, $\phi_{WW}$ is the indication of robustness; on the other end, $\phi_{FF}$ is the indication of resilience. The two states depict dualism. The transitions from $W$ to $F$ and $F$ to $W$ depict interactionism.}

\textcolor{black}{This toy example shows two states but it can be expanded to multiple states, describing different stages of the failure and recovery processes. In this way, a finer grained process of resilience can be depicted. }

This toy example can be further expanded into more sophisticated systems by incorporating many factors. For example, there can be uncertainties associated with the set $\mathcal{A}$, which models the unknown events. In addition, each mode can be associated with a dynamical system, which extends the toy example into a two-mode hybrid system. For example, the underlying controlled system is 
\begin{equation}
\dot{x}=f(x,u, \mathcal{E}_\theta, \theta),
\end{equation}\label{ds}

\noindent where $x\in \mathcal{X}$ is the state; $u\in \mathcal{X}$ is the control that can designed dependent on the mode and the state; $\mathcal{E}_\theta\subseteq \mathcal{E}$ is the set of events considered in mode $\theta\in\Theta$; and $f$ describes the evolution of the dynamics that are dependent on $\theta\in \Theta$.


A control law under complete state information $\mu: \mathcal{X}\times\Theta\rightarrow \mathcal{U}$ and takes the form of $u=\mu(x, \theta)$. The control law depends on the mode of the operation. For a two-mode system, it consists of two components. One designs robust control for the set of inelastic events while the other one designs a resilient control for the set of elastic events.  The formulation of this control law has been extensively examined in the work of Zhu and Ba\c{s}ar \cite{zhu2015game}, utilizing a game-theoretic approach, a perspective that we will delve into further within this article. Additionally, various models have been explored, tailored to diverse attack scenarios within the same framework in \cite{zhu2020cross}.


\section{Robustness-Resilience Contextual Dualism}
Robustness and resilience are two system properties focused on different aspects of a system. They are not antagonistic but distinct in their objectives and capabilities.  The multi-stage illustration provides a temporal dichotomy to distinguish and connect them. Robustness represents a proactive system feature geared towards prevention, whereas resilience constitutes a reactive function focused on response and recovery after an event occurs.  This duality is also shaped by the events that influence the design of a functional and efficient system. It is important to note that this dualism is context-dependent, as it hinges on the specific events under consideration. In essence, the contextual duality arises from the pivotal role that events play in determining whether robustness or resilience should be prioritized.

We typically prepare for robustness in anticipation of events that are foreseeable while the events that we need to prepare for resilience can be either anticipated or unanticipated. Events that fall into the category of expected resilience scenarios are typically those that are challenging to address through robustness measures, often due to their low probability or high prevention costs. Unforeseen events are those that lie beyond the system's prior knowledge. In such cases, resilience becomes the only option to strengthen the system.

Network design is a ubiquitous challenge, often characterized by the nuanced interplay of robustness and resilience, which has led to potential ambiguities surrounding these concepts. When addressing network issues, our focus often revolves around assessing the consequences of link and node failures and devising strategies to safeguard networks against such vulnerabilities. Consider a network's primary objective: to sustain connectivity, ensuring that all nodes within it can effectively communicate with one another. The robustness of a network is measured by its capacity to maintain this connectivity in the face of anticipated failure scenarios, such as random node or link failures. Achieving robustness involves allocating resources to nodes or links, for instance, by introducing redundant nodes or links. This proactive approach ensures that the network remains connected despite these failures.

However, as the scale of potential failures, to be denoted by $n$, increases, preparing for all conceivable $n$-node or $n$-link failures becomes an impractical endeavor. Hence resilience comes to the forefront. Here, the network readies itself to confront unforeseen events, drawing upon a repertoire of resources strategically assembled from various sources, including inherent redundancies and the means to rebuild nodes and links. Crucially, it is evident that robustness serves as a precursor to resilience. The resources invested in enhancing robustness can be strategically repurposed to achieve resilience in subsequent stages, emphasizing the interconnected nature of these concepts in the context of network design.

One illustrative example is the infrastructure network recovery problem that was considered in \cite{chen2019dynamic}. A designer with a budget constraint can spend the resources to build fortified links ahead of time or spend the effort on recovering the links after a disruptive event. Optimal decision depends on the expected severity of the events. It has been observed that if the disruption is calamitous, it is optimal to build a minimum network prior to the disaster and focus the resources on the recovery process to rebuild and augment the network after the disaster. On the contrary, if the event is not highly disruptive, it is optimal for the network to spend most of the resources to build a robust and functioning network at the beginning and spend the remaining resources to fix the links if they are broken after the event. In this way, the infrastructure network can provide a long term service from the very beginning benefiting the crowd.

\section{Robustness-Resilience Interactionism}
The contextual dualism is also accompanied by the interactionism. As seen in the illustrative example above, robustness contributes to resilience while resilience can mitigate events that cannot be handled through robustness. These two system properties work in tandem, fortifying the system against a wide range of events. Their interdependence fosters a symbiotic relationship that intricately weaves them together within a well-designed system. Their designs should not be in silos but instead follow a symbiotic, iterative, and coordinated approach.
 On the one hand,  a resilient design that is built on top of a given robust design fortifies the system by taking care of uncertainties that fall outside the scope of robustness. This scenario is especially pertinent in contemporary systems that require enhanced protection against ever-evolving, sophisticated threats.   On  the other hand, a robust design must incorporate considerations for resilient capabilities and align its level of robustness accordingly.  We can envision robust and resilient designs as two distinct entities engaging in concurrent design processes.  This dynamic can be viewed as  a continuous adaptation, akin to a game-theoretic best response paradigm, where two agents interact, either pursuing common or separate objectives.  


\section{Game-Theoretic Paradigms}

The contextual dualism and interactionism naturally lead to a game-theoretic perspective toward robustness and resilience when they are viewed as two interdependent processes. Furthermore, their interactions with uncertainties can also be captured by a game-theoretic view.
Norbert Wiener introduced the concepts of Manichean devils and Augustinian devils in his work on cybernetics \cite{wiener1988human}, where he sought to elucidate the challenges of uncertainties in science and engineering.   Manichean devils are individuals who employ cunning and dissimulation to outmaneuver the system, often resorting to deceptive and strategic maneuvering to secure victory. They can be likened to chess opponents who operate stealthily and employ deception as their modus operandi. On the other hand, the Augustinian devil is characterized by one that is governed by chance and disorder but still obeys the rules. This type of adversary can even be represented by the forces of nature itself. In this context, nature is not intentionally attempting to outsmart us, but rather, there exist inherent uncertainties within systems that are comprehensible to us.

There has been extensive literature on robustness of control systems to natural uncertainties and disturbances. Within this body of work, game-theoretic models have emerged as a valuable tool for formalizing the concept of robustness (see \cite{bacsar2008h}). This formalization revolves around conceiving robust control as a zero-sum game between the controller and nature's uncertainties. In essence, designing for robustness is a process of playing a game with Augustinian devils. These adversaries operate within predefined behavioral bounds, yet they strive to orchestrate the most adverse outcomes. Robustness, in this context, stands as a guard against the worst-case scenarios that adversaries can engineer, respecting also their limitations.

Manichean devils are more challenging to deal with. Such adversaries possess the ability to be deceptive, strategic, and often remain unknown to the system. Managing these adversarial dynamics calls for resilience. Resilience, in this sense, can be likened to engaging in a nonzero-sum game where the adversaries pursue their objectives clandestinely, concealed from the system's awareness. Meanwhile, the system endeavors to pursue its own objectives, adapting and responding effectively to the maneuvers of these undisclosed adversaries.
Several game-theoretic models have been developed to advance resilient control designs. One notable example of such innovation can be found in \cite{chen2019dynamic}, where a sophisticated hybrid dynamic game model has been introduced. This model serves as providing a comprehensive framework for encapsulating the essence of resilient control.  The dynamic game, as elucidated in \cite{chen2019dynamic}, unfolds in two distinct phases: an {\it ex ante} phase and an {\it ex post} phase, each characterized by different operational modes. Within each of these modes, the control system engages in diverse games against a range of adversaries. This approach provides a holistic and integrated perspective from the lens of game theory for designing control systems that are not only robust but also resilient.

The specific forms that these games take depends on the particular adversary models under consideration. For instance, there are jamming games tailored to scenarios where a system operates in the presence of a jammer \cite{wu2011anti,basar1983gaussian}, a disruptive force capable of interrupting communication channels. Network security games \cite{manshaei2013game,alpcan2010network}, on the other hand, come into play when adversaries seek to control nodes or links within a network to disrupt its operation. A relatively recent addition to this repertoire is deception games \cite{pawlick2019game,pawlick2021game,sayin2018dynamic,sayin2020persuasion}, a novel class of games wherein adversaries adeptly conceal themselves and craft misleading information to manipulate the beliefs and decision-making processes of the systems involved.

\section{Research Directions}

\subsection{Toward a science of resilience}
Robust control theory, which had a fast growth and development in the 1980s and 1990s, has played a pivotal role in establishing a theoretical framework for assessing the robustness of dynamical systems and designing modern control systems and automation solutions. Building upon the success of robust control theory, there arises a similar imperative for a comprehensive theory of resilience. Such a theory would serve as the underpinning for the emergence of resilient control theory and the field of cyber resilience.

The initial step in this endeavor entails the quantification and definition of resilience. Resilience is a universal system attribute that has been investigated across diverse domains such as psychology, business, and cybersecurity.  While its definition must be adapted to suit various types of systems, it should still encapsulate the core essence of resilience. The subsequent stride involves an exploration of the metrics and characterizations associated with resilience properties, as well as the inherent tradeoffs involved.  Figure 1 inherently hints at several prospective metrics. For instance, one could quantify the total performance loss spanning from the start of the event to the end of the incident. Alternatively, a metric could focus on the time required to restore the system from the moment the event happens. By employing suitable metrics, resilience evolves from being solely a qualitative concept into a quantifiable tool for design.

As previously discussed, resilient design is intricately linked to the design of event sets. In a notable study \cite{liu2020robust}, an optimization problem has been formulated, which partitions uncertainty sets into those taken care of by robustness or worst-case measures, and the extreme uncertainties through risk-sensitive risk measures, which inherently call for resilience measures as a means of mitigation. This approach provides an initial foundation for developing more sophisticated designs.

The methodologies for design can be quantified and formalized through the lens of game-theoretic frameworks. These perspectives empower us to quantify both robustness and resilience, extending beyond linear systems. For instance, zero-sum game frameworks have been instrumental in defining and designing robustness, and connections to optimal robust designs under risk-sensitive objective functions have been well established \cite{bacsar2021robust}. Dynamic games have been adept at capturing resilient control design under various types of adversaries. In \cite{huang2020dynamic}, we have discussed the need for dynamic games for resilient control systems and presented several case studies ranging from cyber-physical systems, computer networks, and cyber insurance. As the capabilities of the adversaries vary, different games are chosen to capture the adversarial interactions between the controlled system and the adversary. Dynamic games have provided essentially a dynamic view of the interactions and consequently led to an attack-aware dynamic defense that can balance between the {\it ex ante} prevention (robustness) strategies and {\it ex poste} recovery (resilience) strategies.

The complexity of contextual dualism and interactionism intensifies when systems or subsystems interact with one another. This underscores the necessity for a system-of-systems level understanding of resilience and robustness. Interdependencies among systems can magnify failures, cascading through the network of interconnected systems. Therefore, it becomes imperative to fortify the robustness of one set of systems while relying on the resilience of others. Optimal design at the system-of-systems level takes on increasing significance, particularly in the realm of large-scale multi-agent systems. It is crucial to recognize that the resilience of individual subsystems does not equate to the resilience of the entire system. Consequently, a scientific approach is indispensable in the pursuit of compositional resilience within subsystems.

\subsection{Mechanisms to achieve resilience}
Resilience encompasses a diverse array of mechanisms, offering fertile ground for exploration. Resilience can be attained by leveraging resources both within and outside the system. Common strategies entail the implementation of feedback loops and predictive intelligence. In the contemporary landscape, these strategies benefit immensely from the integration and augmentation of AI techniques, including data analytics, machine learning, generative models, and game-theoretic methodologies. For example, \cite{zhao2022multi} explores multi-agent learning methods within distributed large-scale control systems and outlines the development of an AI stack to offer computational intelligence for subsystems to detect, respond to, and recover from disruptions. Additionally, \cite{xie2020review} presents a comprehensive review of machine learning applications in power systems, focusing on the integration of advanced machine learning technologies with vast amounts of real-time data.

Furthermore, non-technical solutions, such as cyber insurance and prudent investment planning \cite{zhang2019mathtt}, can provide an additional layer of resilience to control systems. These measures ensure that system owners are provided sufficient financial resources for effective system recovery in the event of malfunctions. Specifically, cyber insurance offers crucial financial coverage, effectively transferring unmitigated risks. This is particularly vital for small business owners who may lack the full capacity to facilitate system restoration following a cyberattack.

\section{Relevance to Control Systems Community}

\textcolor{black}{Our society grapples with the ever-growing challenges posed by global risks such as climate change,  conflicts, and cyber threats. In the face of these uncertainties, there is a growing demand for engineering system designs that exhibit high resiliency. The field of resilient control and engineering emerges as a crucial player in safeguarding our critical infrastructure, securing cyber systems, and ensuring high assurance in interactive human-engineering systems. For instance,  our critical infrastructures, including energy systems, transportation networks, and water systems, are not only vulnerable to failures caused by natural events but are also increasingly susceptible to intentional adversarial attacks, in both the physical system and the cyber space. Compounding this vulnerability is the intricate interconnection of our infrastructures; a failure in one critical system can cascade into further failures across others. The imperative to develop resilient solutions to mitigate these risks becomes evident. To this end, methodologies rooted in control, network, and game theory, as explored in \cite{chen2019game,rass2020critical}, are promising ones as they provide a holistic understanding of the interdependent infrastructure networks and incorporate uncertainties that were traditionally excluded from modeling frameworks. }



\textcolor{black}{In the cyberspace, resilience is crucial for enabling computer networks and cyber operations to anticipate, respond to, and recover from cyber incidents. Cyber resilience serves as a vital tool for organizations grappling with ransomware, insider threats, and zero-day attacks, allowing them to mitigate risks that may not have been experienced in the past. As we approach 2030, these risks will escalate and become more critical as they are hard to eliminate. Control and game theory stand out as providing indispensable tools within which to address  this challenge at a foundational level. The conflicting goals of attackers and defenders find a natural representation in game-theoretic frameworks \cite{alpcan2010network}. The feedback nature of the ``anticipate," ``respond," and ``recover" loop can be captured using a control feedback system. As elaborated in \cite{huang2022reinforcement}, the synergy between game theory, control theory, and learning theory paves a promising way for a scientific theoretical foundation for cyber resilience. Designing cyber-resilient mechanisms, incorporating cyber insurance \cite{zhang2019mathtt}, zero trust \cite{ge2023gazeta}, and cyber deception \cite{sayin2018dynamic}, will be a critical and promising challenge for the community to address as we welcome the next-generation engineering systems by 2030, such as the quantum information technologies, 5G-enabled autonomous systems, and large-scale cyber-physical systems and networks. }

\textcolor{black}{In addition to the cyber and physical domains, human behaviors are often regarded as among the most challenging aspects of engineering systems. Human behaviors can lead to unexpected outcomes in optimally designed engineering systems, as evidenced by phenomena such as the Braess paradox \cite{nagurney2007internet} and the ostrich paradox \cite{meyer2017ostrich}. Moreover, human errors and vulnerabilities can cause system failures and disruptions that engineering systems, designed to be robust against physical modeling errors and perturbations, may not anticipate. Noted in \cite{huang2023cognitive}, attackers can exploit human vulnerabilities through tactics like misinformation \cite{bastopcu2023role,kapsikar2020controlling}, phishing \cite{huang2022advert}, and gaslighting \cite{liu2023information}, resulting in intended disruptions to cyber-physical systems.  Addressing the risks posed by human behaviors through the incorporation of resilience into control system designs presents a promising research avenue for next-generation control scientists and engineers. One effective approach is to integrate game theory into the control design process, given its role as a fundamental framework for modeling and explaining human incentives and rational behaviors. Furthermore, recent developments in behavioral game theory have explored models and frameworks capturing agents with bounded rationality \cite{camerer2011behavioral}, and incorporating these advances into control theory holds promise for establishing a foundation to design resilient mechanisms against human vulnerabilities. Looking ahead, the next decade will witness an increase in networks and systems serving and interacting with human users. This trend aligns well with the growing popularity of on-demand services, social media, virtual and augmented reality. Resilience emerges as an indispensable component, ensuring the dependability of these systems and products in the face of evolving human behaviors and interactions.}

\textcolor{black}{Moreover, resilience is the key to sustainability challenges that we face today as it enables systems to withstand and rebound from various challenges and disruptions. By building resilience, we ensure that ecosystems, social structures, and economic systems can adapt to changing conditions, reducing the risk of irreparable damage and promoting long-term sustainability. Resilient systems can better cope with environmental threats such as climate change and natural disasters, as well as social and economic shocks, thereby enhancing their ability to persist and thrive over time. The concept of resilience extends beyond adversarial behaviors  to other disruptive events such as climate changes \cite{bernhardt2013resilience,bjork2008managing}, geopolitical conflicts \cite{alqahtani2021oil,carvalho2014resilience}, financial crises \cite{scoblic2020emerging,markman2014resilience}, and natural disasters \cite{he2018robust,zimmerman2017conceptual}. Determining whether a natural disaster event should be classified as an inelastic or elastic event is a strategic design choice, particularly if it falls within the overlap of these categories. In such cases, a combination of methods from both robust control and resilient control can be deployed to mitigate their impact. This reflects the interactionism of the dualism discussed  in the article. Understanding budget constraints and system-domain specifications plays an important role in understanding  the tradeoffs between robustness and resilience. Design considerations for events vary across different systems. For example, mission-critical military systems face distinct concerns compared to the events that can affect household heating control systems. These domain-specific design processes lead to different outcomes.}

\textcolor{black}{The control community has a rich history of developing solutions to deal with disturbances, uncertainties, and disruptions. We have witnessed the success of robust control theory, which aims to attain robust performance and stability with bounded uncertainties. Moving toward 2030, we find ourselves at the right time to address challenges associated with resilience. It is an opportune moment for the community to lead the charge in establishing the scientific foundation of resilience. Achieving this goal relies on  the multifaceted expertise within our community, encompassing feedback control theory, game theory, and learning theory. The science and engineering of resilience is inherently transdisciplinary as we deal with systems with growing scale and complexity and risks from diverse domains.  Collaborative efforts within our society, and across diverse societies, will serve as driving forces  in establishing the scientific foundations and advancing the development of next-generation control systems and applications. This collective endeavor holds the potential for significant societal impact, with improved quality of life evident when our infrastructure systems showcase resilience to natural disasters like flooding and hurricanes caused by climate change. Resilient control systems will contribute to a growing economy by withstanding geopolitical disruptions in supply chains, overcoming shortages in food systems, thwarting cyber threats to energy systems, and maintaining democratic resilience against misinformation and disinformation. } 
%

\newpage

\section{Box I: Dualism and Interactionism}

\textcolor{black}{Dualism between  resilience and robustness propounds that robustness and resilience are fundamentally distinct and cannot be reduced to each other. Although of distinct natures, they interact with each other. }

\textcolor{black}{Interactionism propounds that robustness and resilience are distinct system properties that interact with each other. This interactionism is contextual and it can vary depending on the system's application domain, design objectives, and economic constraints. This contextual interactionism explains the relationship between robustness and resilience within their dualism. Recognizing these contextual differences is crucial to understanding how robustness and resilience can work together to achieve control system objectives. }

\newpage
\section{Box II: Elastic and Inelastic Events}

\textcolor{black}{For a given system, there are elastic events naturally addressed through resilience, and inelastic ones best tackled through robustness. However, some events fall into a nuanced middle ground, with their classification as elastic or inelastic being contingent upon the system's design objectives.  Consider, for instance, an energy system designed to exhibit robustness against events resulting in the failure of a single node, in accordance with the $N-1$ criterion. Yet, if the system is required to endure events causing a greater number of node failures, prioritizing resilience becomes the appropriate choice. In scenarios involving mission-critical energy systems, the decision may lean towards robustness against  failures, despite the associated increase in design costs. Consequently, events leading to two-node failures are designated as inelastic events.  This classification of events into elastic and inelastic sets represents thoughtful design choices. It strikes a balance between costs and benefits while meeting the system specifications. 
Elasticity offers one dimension to categorize events into ones that need to be handled through robustness and the ones that can be handled through resilience. There are other dimensions that can be introduced into the decision when we need to make distinctions among fault-tolerance, robustness, and resilience. Readers can refer to \cite{zhao2022multi} for an explanation on the distinction between fault-tolerance and resilience, which goes beyond the scope of this article.}

\begin{figure}[hb]
    \centering
    \includegraphics[width=0.6\textwidth]{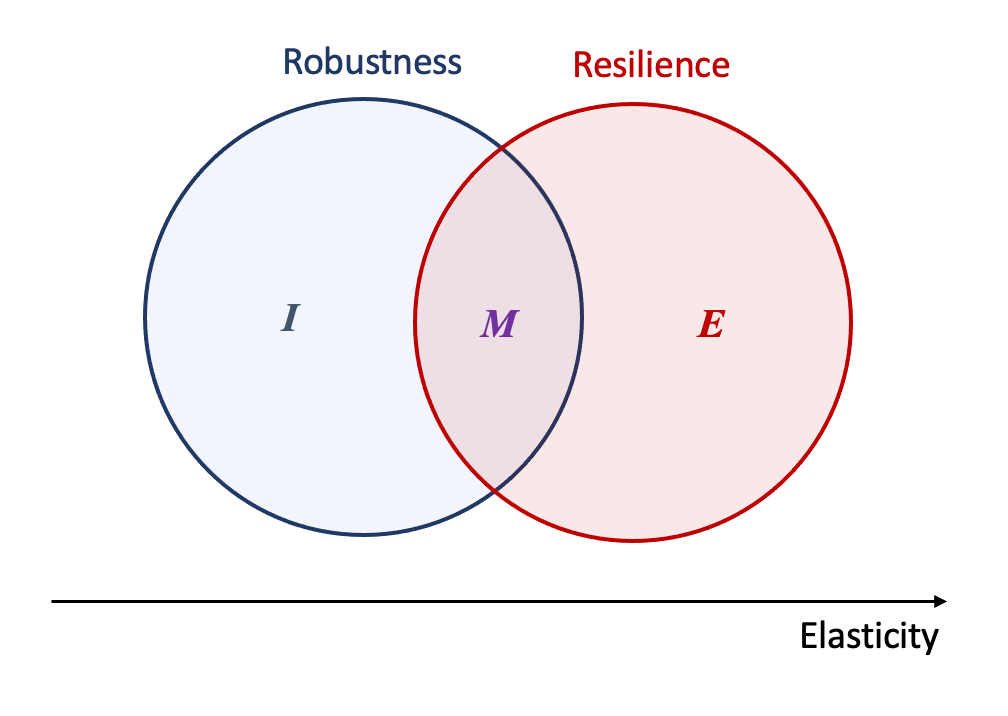}
    \caption{\textcolor{black}{Events are classified into two categories based on their elasticity: the set of elastic events, denoted by $E$,  and the set of inelastic events, denoted by $I$. The set of events in the middle, denoted by $M$, can be managed through either robustness or resilience strategies. The choice between these strategies is determined by system specifications or a cost-and-benefit analysis.}
}
    \label{fig:events}
\end{figure}

\newpage

\bibliographystyle{IEEEtran}
\bibliography{reference}

\begin{thebibliography}{10}
\bibitem{markman2014resilience}
G.~M. Markman and M.~Venzin, ``Resilience: Lessons from banks that have braved
  the economic crisis—and from those that have not,'' \emph{International
  Business Review}, vol.~23, no.~6, pp. 1096--1107, 2014.
\end{thebibliography}

\end{document}